\documentclass[referee]{cjaa}           

\usepackage{graphicx}                   
\input{epsf.sty}                        
\input{psfig.sty}                       

\newcommand{\be}{\begin{equation}}
\newcommand{\ee}{\end{equation}}

\begin{document}

   \title{The X-ray emission lines in GRB afterglows: the evidence for the
   two-component jet model
}

   \volnopage{Vol.0 (200x) No.0, 000--000}      
   \setcounter{page}{1}          

   \author{W.H. Gao\inst{1,2}\mailto{whgao@pmo.ac.cn} \and D.M. Wei\inst{1,2}
}
   \offprints{D.M. Wei}                   

   \institute{Purple Mountain Observatory, Chinese Academy of
Sciences, Nanjing, 210008, China\\
             \email{whgao@pmo.ac.cn}
         \and National Astronomical
Observatories, Chinese Academy of Sciences, Beijing, 100012,
China\\
}
   \date{Received~~2005 month day; accepted~~2005~~month day}

\abstract{Recently, X-ray emission lines have been observed in
X-ray afterglows of several $\gamma$-ray bursts. It is a major
breakthrough for understanding the nature of the progenitors. It
is proposed that the X-ray emission lines can be well explained by
the Geometry-Dominated models, but in these models the
illuminating angle is much larger than that of the collimated jet
of the $\gamma$-ray bursts(GRBs). For GRB 011211, we obtain the
illuminating angle is about $\theta\sim45^{\circ}$, while the
angle of GRB jet is only $3.6^{\circ}$, so we propose that the
outflow of the GRBs with emission lines should have two distinct
components. The wide component illuminates the reprocessing
material, and produces the emission lines, while the narrow one
produces the $\gamma$-ray bursts. The observations show that the
energy for producing the emission lines is higher than that of the
GRBs. In this case, when the wide component dominates the
afterglows, a bump will appear in the GRBs afterglows. For GRB
011211, the emergence time of the bump is less than 0.05 days
after the GRB, it is obviously too early for the observation to
catch it. With the presence of the X-ray emission lines there
should also be a bright emission component between the UV and the
soft X-rays. These features can be tested by the $Swift$ satellite
in the near future. }

   \authorrunning{W.H.Gao \& D.M.Wei}            
   \titlerunning{The X-ray emission lines: the evidence for the
   two-component jet model}  

   \maketitle

%
%
\section{Introduction}           
\label{sect:Intro}
Gamma-ray bursts(GRBs) are commonly interpreted in terms of a
relativistic outflow emanating from the vicinity of a stellar
neutron star or black hole(e.g. Piran 2004; Zhang \&
M$\acute{e}$sz$\acute{a}$ros 2004). Highly collimated narrow jets
can be inferred from the presence of achromatic breaks in the
afterglow lightcurves(Rhoads 1999).

Recently X-ray emission lines have been observed in X-ray
afterglow of several GRBs. They can provide important clues for
identifying the nature of the progenitors of long (t$\geq$2 s)
GRBs. The first marginal detection of an emission line was in the
X-ray afterglow of GRB 970508 with the BeppoSAX NFI (Piro et al.
1999). Later emission lines were also detected in the X-ray
afterglows of GRB 970828 (Yoshida et al. 2001) with ASCA; GRB
991216 (Piro et al. 2000) and GRB020813 (Butler et al. 2003) with
Chandra; GRB 011211 (Reeves et al. 2002), GRB 001025A (Watson et
al. 2002) and GRB 030227 (Watson et al. 2003) with XMM-Newton; GRB
000214 (Antonelli et al. 2000) with BeppoSAX. The detailed
properties of the X-ray emission features can be found in several
papers ( Lazzati 2002; B\"{o}ttcher 2003; Gao \& Wei 2004 ). The
locations of the emission lines found in the X-ray afterglows of
GRB 970508, GRB 970828, GRB 991216 and GRB 000214 are roughly
consistent with Fe $K_{\alpha}$ at the redshift of the hosts,
while the emission lines were identified as blueshifted light
elements lines of S, Si, Ar, Mg, and Ca in the afterglow of GRB
011211, GRB 020813 and GRB 030227.

Two main types of models have been put forward to interpret the
emission lines: one is Geometry-Dominated(GD) models(e.g. Vietri
et al. 2001; Lazzati et al. 1999; Reeves et al. 2002), the other
is Engine-Dominated(ED) models(e.g. Rees \&
M$\acute{e}$sz$\acute{a}$ros 2000; M$\acute{e}$sz$\acute{a}$ros \&
Rees 2001 ). In ED models, the lines are created by reprocessing
material very close to the explosion site(R$\sim 10^{13}cm$). The
ionizing continuum is believed to be provided by a post-burst
energy injection(Rees \& M$\acute{e}$sz$\acute{a}$ros 2000;
M$\acute{e}$sz$\acute{a}$ros \& Rees 2001; Gao \& Wei 2004, 2005).
The duration of the lines emission is determined by the time
interval of the post-burst energy injection. While in GD models
the reprocessing material is located at a large enough
distance($R\sim 10^{16}cm$), illuminated by the burst and early
afterglow photons. In GD models, the duration of the emission
lines is set by the size of the reprocessor. This reprocessing
material is compact and metal enriched, similar to a supernova
remnant, as predicted in the supranova model(Vietri \& Stella
1998).

It is argued that the production of the emission lines strongly
favors the GD models(e.g. Lazzati et al. 2002; Reeves et al.
2002). But in these models, either in the reflection model or in
the thermal model, a large collimation angle of the illuminator is
needed(e.g. Lazzati et al. 1999, Reeves et al. 2002).

The half opening angle of the GRB jet is obtained from the
presence of the achromatic break in the afterglow lightcurve(Frail
et al. 2001, Bloom et al. 2003), it is much smaller than the
illuminating angle obtained with the GD models. If the photons of
illuminating the reprocessing material comes from the bursts and
early afterglows, with such small collimation angle, the duration
time of the emission lines will be much shorter than that of the
observations. It also contravenes the fact that much higher energy
is needed for the illuminating continuum that is responsible for
the lines production than that of the collimated GRBs(Lazzati
2002; Ghisellini et al. 2002; Gao \& Wei 2004). To solve the
energy problem, we have explained it as continuous post-bust
energy injection from magnetar(Gao \& Wei 2004), or delayed energy
injection from central engine(Gao \& Wei 2005), similar model can
also explain the early flare in X ray afterglow light curve(Fan \&
Wei 2005).

Recently, two distinct components in the GRB outflow has been
proposed for several gamma-ray bursts. On the observational side,
Frail et al.(2000) proposed that the $\gamma$-rays and early
X-rays and optical afterglow of GRB 991216 could be attributed to
a narrow ultra-relativistic outflow component and the
longer-wavelength afterglow such as radio afterglow originated in
a wide component that is only mildly relativistic. A similar
picture was proposed for GRB 970508(Pedersen et al. 1998) and GRB
030329(Berger et al. 2003; Sheth et al. 2003). A two-component
model was also suggested as the explanation for the observation of
the re-brightening of the X-ray flash source XRF 030723(Huang et
al. 2004). In the numerical simulations of collapsar model, the
narrow component has a Lorentz factor $\gamma_{n}\geq$100 and a
half opening angle $\theta_{n}\sim3^{\circ}-5^{\circ}$, while for
the wide component, $\gamma_{w}\sim$15 and
$\theta_{w}\sim10^{\circ}$(Zhang et al. 2004). In his GRMHD models
of the jet formation, McKinney(2005) has found a two-component jet
with a quite broad component out to $25^{\circ}$ and a core
component within $5^{\circ}$.

In this paper, we reconsider the Geometry-Dominated models, and
investigate the angle of the illuminator in GRB 011211 afterglow.
A large illuminating angle is obtained. Here we propose a
two-component outflow model to solve the angle problem. We propose
that the outflow in the GRB whose afterglow shows the lines
emission has two components, the angle of the wide component is
about several tens degrees, and the energy of the wide component
is high enough to illuminate the material to produce the emission
lines.


\section{The illuminating angle and the two-component jet model}
\label{sect:Model}
In this paper the Geometry-Dominated models are used to explain
the lines production(e.g. Lazzati 2002, Lazzati et al. 2002). In
these models the lines emission comes from an extended region and
its duration arises from light-travel time effects. So the
geometry of the dense material is important because it dominates
the time duration of the line emission. Lazzati et al.(1999) have
investigated the relations between the geometry of the
reprocessing material and the theoretical models.

Here we investigate the illuminating angle in the reflection model
and thermal model. In the reflection model, the photons are
reflected by the funnel-shaped material. Here we adopt the
geometry of the reprocessing material similar to that adopted in
the work of Tavecchio et al.(2004). This geometry of the material
is efficient for the reflection, we called it funnel-shaped
reprocessing material(Fig.1a). In the first subsection below, we
re-calculate the illuminating angle in the reflection model. The
detailed calculation of the physical quantities can be found in
the paper of Tavecchio et al.(2004).

In the thermal model, similar to the geometry taken by Lazzati et
al. (1999), we adopt the shell-shaped reprocessing
material(Fig.1b). This geometry of the material is efficient to be
heated to high temperature and produces the emission lines by
collision-ionization and recombination. Though Reeves et al.(2002)
have explained the production of the emission lines with the
thermal model, an illuminating angle of $20^{\circ}$ is only
assumed(not calculated). In the second subsection below we
calculate the illuminating angle of the burst source in the
thermal model.

\begin{figure}
   \begin{center}
   \epsscale{1.0}{1.0}
   \includegraphics[angle=0,width=12cm]{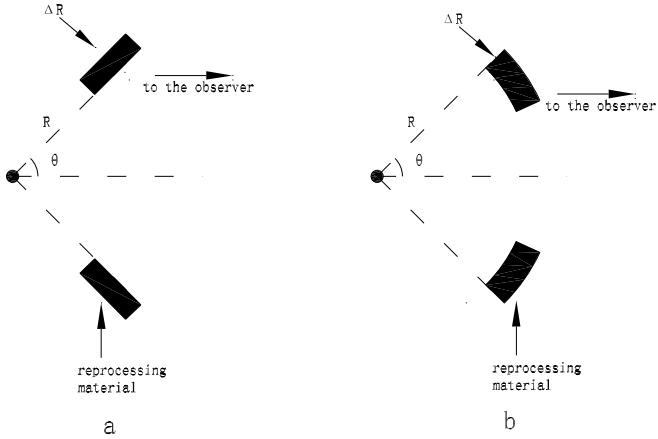}
   \caption{Sketch of the geometry assumed for the reprocessing
material in the Geometry-Dominated models. a: the funnel-shaped
material in reflection model. b: the shell-shaped material in
thermal model.}
   \label{Fig:fig1}
   \end{center}
\end{figure}

\subsection{The illuminating angle in the reflection model}

Here we reconsider the reflection model, similar to the
calculation of Tavecchio et al.(2004). In the reflection model,
with the funnel-shaped reprocessing material(Fig.1a), the time at
which lines become visible can be defined as following(the time
intervals are measured in the frame of the burst source):
\begin{equation}
t_{app}=\frac{R}{c}(1-cos\theta),
\end{equation}
where $\theta$ is the half opening angle of the illuminator, R is
the distance between the burst source and the reprocessing
material.

The duration of the line $t_{line}$ can be given by
\begin{equation}
t_{line}\sim\frac{\Delta R}{c}cos\theta,
\end{equation}
where $\Delta R$ is the width of the reprocessing material, which
can be expressed as (Tavecchio et al. 2004).
\begin{equation}
\frac{\Delta
R}{R}=\frac{1}{5\pi}\xi\zeta\frac{W_{i}m_{p}}{A_{i}\epsilon_{i}sin\theta\alpha_{r}},
\end{equation}
where $W_{i}$ is the atomic weight of the element(the subscript i
denotes a specific line). $\xi$ is the ionization parameter. For
soft X-ray lines, $\xi$ is about $10^{2}$ for the efficient
emission(Lazzati et al. 2002). For GRB011211, ionization parameter
is
\begin{equation}
\xi=\frac{L_{ill}}{n_{e}R^{2}}\simeq10^{2},
\end{equation}

here $L_{ill}$ is the luminosity of the X-ray illuminating
continuum. $n_{e}$ is the number density of the electron. $\zeta$
is the efficiency in producing the lines(e.g. Ghisellini et al.
2002),

\begin{equation}
\zeta=\frac{E_{line}}{E_{ill}}\simeq10^{-2}.
\end{equation}

$A_{i}$ is the mass abundance of the emitting element,
$\epsilon_{i}$ is the energy of the line, $\alpha_{r}$ is

\begin{equation}
\alpha_{r}=5.2\times10^{-14}Z\lambda^{1/2}[0.429+0.5ln(\lambda)+\frac{0.496}{\lambda}^{1/3}],
\end{equation}

where $\lambda=1.58\times10^{5}Z^{2}T^{-1}$, Z is the atomic
number of the element and T is the electron temperature. For GRB
011211, the temperature T of a photoionized plasma illuminated
with $\xi=100$ is predicted to be in the range
$10^{5}-10^{6}$K(e.g. Kallman \& McCray 1982). Here we then
assumed $T=5\times 10^{5}$K.

Combing Eq. (1)-(3), it can be found that

\begin{equation}
tan\theta(1-cos\theta)\sim\frac{1}{5\pi}\xi\zeta\frac{W_{i}m_{p}}{A_{i}\epsilon_{i}\alpha_{r}}\frac{t_{app}}{t_{line}}.
\end{equation}

For GRB 011211, the observation shows that the emission lines
appear at time $t_{app}\leq4\times10^{4}/(1+z)s
\sim1.3\times10^{4}s$, and the time duration of the lines
$t_{line}\simeq5\times10^{3}/(1+z) s\sim1.7\times10^{3}s$.
$\xi\simeq100$ and $\zeta\simeq0.01$. For the solar abundance of
the elements(Anders \& Grevesse 1989), we can know the value of
$A_{i}$. $\epsilon_{i}$ is the center energy of the line. For
instance, $\epsilon_{CaXX}$ is 4.70 KeV(in the burst source
frame). In this case, the illuminating angle can be obtained about
$\theta\sim45^{\circ}$.

\subsection{The angle of the illuminator in the thermal model}

In the thermal model, for the shell-shaped reprocessing
material(Fig.1b), the duration of the lines observed in the GRB
afterglow is
\begin{equation}
t_{line}=\frac{R}{c}(1-cos\theta).
\end{equation}

The X-ray lines luminosity is
\begin{equation}
L_{i}=[N_{i}\epsilon_{i}/t_{rec}](1+z)^{-1},
\end{equation}
where $N_{i}$ is the number of specific element nuclei. $t_{rec}$
is the recombination time scale. When the reprocessing material is
heated to thermal equilibrium, close to the Compton temperature
$T_{C}\sim10^{7}K$(Reeves et al. 2002), we can get that
$t_{rec}\sim10^{11}T_{7}^{1/2}n_{e}^{-1}$.

The number of specific element nuclei in the material layer within
Thomson optical depth $\tau_{T}$=1(in $\tau=1$ layer, the material
absorbs enough energy without smearing the lines very much(see Gao
\& Wei 2005)) is
\begin{equation}
N_{i}\sim A_{i} M/(\tau_{T}W_{i} m_{p}),
\end{equation}
where M is the total mass of the material that is illuminated by
the illuminator. $M=n_{e}m_{p}V$, V is the volume of the
illuminated material, and $V=2\pi R^{2}(1-cos\theta)\Delta R$.
$\tau_{T}$ is Thomson optical depth of the reprocessing material,
$\tau_{T}=n_{e}\Delta R \sigma_{T}$.

For GRB 011211, the ionization parameter $\xi\simeq10^{2}$(Lazzati
et al. 2002), the X-ray luminosity illuminating the material is of
the order $L_{ill}\sim 10^{47}erg s^{-1}$. The luminosity of the
line is about $10^{45}erg s^{-1}$, $t_{line}\sim1.7\times10^{3}s$.
According to these, we can get $\theta\sim 45^{\circ}$.

We have obtained the angle of the illuminator, that is about
$45^{\circ}$. Obviously it is much larger than the half opening
angle of the GRB jet, that is only $3.6^{\circ}$(Frail et al.
2001; Bloom et al. 2003).

Therefore, we propose that the outflow of the GRB with the X-ray
lines has two distinct components: the narrow one produces the
prompt GRB emission, while the wide component illuminates the
reprocessing material and produces the emission lines. The
illuminator's angle is $\theta\sim\theta_{w}$.

\section{Production of the bump in two-component outflow}

It has been found that the energy obtained from the lines emission
is higher than that of the $\gamma$-ray burst(Ghisellini et al.
2002; Gao and Wei 2004). So in two-component outflow model, the
energy of the wide component illuminating the reprocessing
material should be $E_{ill}\sim E_{w}\geq E_{n}$, $E_{n}$ is the
energy of the narrow component, that is the same as that of the
$\gamma$-ray burst, $E_{n}=E_{\gamma}$.

In the theory of GRB afterglow, the interaction of the jet with
the ambient medium drives a reverse shock into the GRB ejecta,
which decelerates the ejecta. The energy given to the swept-up
external medium is about $\sim\eta^{2}M_{sw}c^{2}$(Here, $\eta$ is
the Lorentz factor of the outflow; $M_{sw}$ is the rest mass of
the swept-up external medium). We assume that the Lorentz factor
of the GRB jet(i.e. the narrow component) is $\eta_{n} \sim 300$.
The energy in the wide component is more than that of the narrow
component, $E_{w}\geq E_{n}$. We assume that the density of the
medium in the wide component is the same with that in the narrow
component. So we can assume $\eta_{w} \geq 30$($\eta_{w}$ is the
Lorentz factor of the wide component).

From the work of Peng er al.(2005), for $E_{w}/E_{n}>1$, we expect
that a bump should be found in the afterglow at the time
$t_{dec,w}$(the deceleration time of the wide component) when the
wide component dominates the afterglow. That is

\begin{equation}
t_{dec,w}\leq0.05(\frac{E_{iso,52}}{n_{0}})^{1/3}(\frac{\eta_{w}}{30})^{-8/3}
days.
\end{equation}

So we should observe the bump at about less than 0.05 days after
the $\gamma$-ray burst. The GRB 011211 X-ray afterglow is observed
about 11 hours after the burst(Reeves et al. 2002). Obviously it
is too early to observe the bump.

If the energy of the wide component is less than or comparable
with that of the narrow component, the lightcurves of the
afterglow would not be dominated by the wide component, so the
bump would not appear. Only when the energy of the wide component
is much higher than that of the narrow one, the bump would
distinctly emerge(Peng et al. 2005; Wu et al. 2005).

\section{Discussion and Conclusions }

The Geometry-Dominated models have been proposed to explain the
X-ray emission lines observed in the X-ray afterglows (e.g.
Lazzati et al. 2002, Reeves et al. 2002). In these models, the
time duration is set by the geometry of the reprocessing material.
In this paper, we investigate the Geometry-Dominated models and
calculate the illuminating angle of the illuminator.

Generally iron line can be well explained in reflection model
while soft X-ray lines prefer the thermal model. Of course, this
is not absolute. Tavecchio et al.(2004) have claimed that emission
lines in GRB 011211 afterglow also can be well explained in
reflection model. So in this paper, we calculate the illuminating
angle in reflection model and thermal model respectively.

For GRB 011211, in the GD models, an illuminating angle
$\theta\sim 45^{\circ}$ is obtained. However from the presence of
the break in the light curve of GRB afterglow, the jet of GRB
011211 with angle only $\theta_{j}\sim 3.6^{\circ}$ has been
obtained(Frail et al. 2001; Bloom et al. 2003).

Since it is comparable for the physics parameter of X-ray lines on
the whole(Lazzati 2002), we assume that the illuminating angle in
other GRBs has the same case as in GRB 011211.

It should be noted that solar abundance of the line element is
adopted in our calculation. The angle would be smaller than
$45^{\circ}$ if we adopt lager element abundance. For instance, if
we adopt 10 times solar abundance of the line element, we can get
$\theta\sim22^{\circ}$. But in any case it will be lager than the
half opening angle of GRB 011211.

Therefore we proposed two-component outflow model in the
$\gamma$-ray burst with X-ray emission lines. The angle of the
wide component is comparable with that of the illuminator,
$\theta\sim 45^{\circ}$; while the collimated jet of the GRB is
comparable with the narrow component, $\theta_{j}\sim
3.6^{\circ}$.

For GRB 011211 the energy for illuminating the reprocessing
material obtained from the X-ray lines is higher than
$5\times10^{50}$ergs(Gao and Wei 2004; Ghisellini et al. 2002),
which is higher than that of the $\gamma$-ray bursts. In this
case, the wide component will dominate the afterglow, a bump will
appear in the lightcurve of the afterglow. Our calculation shows
that it should be seen at about less than 0.05 days after the
burst in the X-ray afterglow. Unluckily it was too early to be
observed.

In this two-component outflow model, when the wide component
dominates the lightcurve of the afterglow, an approximate
isotropic component of the X-ray afterglow should be observed. In
the X-ray band observation of GRB 011211 afterglow, no break in
the afterglow was reported(Reeves et al. 2002).

For GRB 991216, a narrow half opening beaming angle $\theta_{j}
\sim 3^{\circ}$ has been claimed from the optical
observation(Halpern et al. 2000). But recently Ruffini et
al.(2005) have claimed that the data analysis in the 2-10 KeV of
GRB 991216 afterglow conformed spherical symmetry. We have
calculated the energy of the illuminator obtained by the emission
lines. It is between 3$\times10^{51}$ and
3.8$\times10^{52}$ergs(Gao and Wei 2004). This energy is much
higher than that of the burst. So the bump would appear in the
lightcurve of the afterglow. But it is too early for the
observation to be observed. When the wide component dominates
X-ray afterglow, a spherical symmetry afterglow would be obtained.
So the result obtained by Ruffini et al.(2005) is consistent with
our two-component outflow model.

In the same way, our two-component model is also consistent with
the observation of GRB 970508, whose afterglow could be explained
in terms of a narrow jet surrounded by an isotropic
outflow(Pedersen et al. 1998).

In the low-ionization condition discussed here, about $90\%$ of
the total incident luminosity is absorbed by the reprocessing
material(e.g. Zicky et al. 1994), and that will be re-emitted. The
reprocessed luminosity $L_{repr}$ is about
\begin{equation}
L_{repr}=0.9\frac{L_{line}}{\zeta}\sim 9\times
10^{46}\frac{L_{line,45}}{\zeta_{-2}}ergs^{-1}.
\end{equation}

The surface of the slab illuminated by the incident continuum will
be heated to high temperature, close to the Compton temperature
$T_{C}\sim 10^{7} K$(e.g. Reeves et al. 2002). The emission from
these outer layers will peak in the X-ray region, at energies
around 1 KeV. Gas in the layers deeper in the material will be
close to the thermal equilibrium: the temperature of these region
will be close to the black-body temperature corresponding to a
black-body emission with the luminosity of the order
$L_{bb}=L_{repr}$. Tavecchio et al.(2004) have obtained that the
maximum of the emission falls at the frequency $\nu_{bb}\sim
10^{15}$ Hz, i.e. in the near UV.

The actual outcoming spectrum will be a complex integral over the
emission from the different layers, in time-dependent conditions.
The detailed calculation of the spectrum is beyond the scope of
this paper.

So, accompanying with the presence of the X-ray emission lines
there should be a bright component between the UV and the soft
X-rays, that is 100 times larger than the line
luminosity(Tavecchio et al.2004).

In conclusion we assume that the outflow of GRBs has two
components. The wide component illuminates the reprocessing
material, and produces the emission lines; while the narrow
component is corresponding to the jet of GRBs. We have obtained
that for GRB 011211 the bump should be observed at less than 0.05
days after the burst because of the higher energy of the wide
component, but it was too early for the bump to be observed. A
bright component between the UV and the soft X-rays should also be
observed with the presence of the X-ray emission lines.

\acknowledgements  We thank S.L. Chen very much for his providing
the figure in this paper. We thank the referee for his valuable
comments. This work is supported by the National Nature Science
Foundation(grants 10233010 and 10225314) and the National 973
Project on Fundamental Researches of China(NKBRSF G19990745).

\label{lastpage}

\end{document}